# The electronic transport properties and microstructure of carbon nanofibre/epoxy composites

A. Allaoui, S.V. Hoa[*] and M.D. Pugh

Concordia Centre for Composites, Department of Mechanical and Industrial Engineering, Concordia University, Montréal, Québec, Canada H3G 2W1

## ABSTRACT

Carbon nanofibres (CNF) were dispersed into an epoxy resin using a combination of ultrasonication and mechanical mixing. The electronic transport properties of the resulting composites were investigated by means of impedance spectroscopy. It was found that a very low critical weight fraction ($p_c$ = 0.064 wt %) which may be taken to correspond to the formation of a tunneling conductive network inside the matrix. The insulator-to-conductor transition region spanned about one order of magnitude from 0.1 to 1 wt %. Far from the transition, the conductivity increased by two orders of magnitude. This increase and the low value of the conductivity were explained in terms of the presence of an epoxy film at the contact between CNF. A simple model based on the CNF-CNF contact network inside the matrix was proposed in order to evaluate the thickness of that film.

[*] Corresponding author. Fax: +1 514 848 3175. E-mail address: hoasuon@alcor.concordia.ca (Suong V. Hoa).

## 1. INTRODUCTION

Carbon nanofibres (CNF) are hollow cylinders with diameters around one hundred nanometers and lengths of a few tens of microns giving high aspect ratios (length/diameter > 100). CNF are the cheaper counterpart of carbon nanotubes (about 3 to 500 times cheaper respectively when compared to multi-wall (MWNT) or single-wall carbon nanotube (SWNT)). They have a larger diameter (2 to 100 times respectively compared to MWNT or SWNT) and are less crystalline (with a kind of cup-stacked or stacked coins structure), while keeping acceptable mechanical and physical properties (Young modulus ~500 GPa, tensile strength ~3 GPa, electrical conductivity ~$10^3$ S/cm, thermal conductivity ~1900 $W.m^{-1}.K^{-1}$). They are expected to be a promising nanofiller for the preparation of composites with multiple enhanced properties. Many studies related to the enhancement of the mechanical properties of an epoxy matrix by the introduction of CNF have been conducted [1-6, 10]. Some studies were dedicated to the electrical properties of CNF/epoxy composites [8-13] and a general review of the properties of CNF-based composites has been published [7]. Because of their high aspect ratio and of van der Waals attractive interactions arising at the nanoscale, CNFs are tangled and form aggregates of different sizes which makes their homogeneous dispersion inside the matrix one of the main hurdles. For this reason, most of the research work published so far on the electrical properties of CNF-based composites focused on relatively high loadings, usually higher than 2 wt %, and aimed to obtain a high conductivity without determination of the critical weight fraction at which the system becomes conductive. In the literature, different preparation methods have been used to disperse CNF inside an epoxy resin and they led to different levels of electrical conductivity of the composites. The use of ultrasonication [10,11] gives a higher level of conductivity, up to $1.2 \times 10^{-2}$ S/cm at 10 wt % loading [10], compared to mechanical mixing [8,9] giving rise to a maximum conductivity of $2 \times 10^{-5}$ S/cm at 8 wt % loading [8]. Almost no data were published for CNF loading lower than 1 wt %. In the present paper, a method combining ultrasonication and mechanical mixing was used to prepare composites at different weight fractions, from 0.066 to 9 wt %, and their electrical properties were investigated by impedance spectroscopy. A complete picture of the electrical properties of a CNF/epoxy composite is given, including the vicinity of the insulator-to-conductor transition and the region far from the transition.

## 2. EXPERIMENTAL

### 2.1 Materials

Heat treated graphitized CNF (Pyrograf III PR24LHT) with diameters in the range of 60-150 nm and lengths between 30 and 100 µm, were obtained from ASI (Figure 1). A low viscosity diglycidyl ether of bisphenol-A (DGEBA) epoxy resin ($\eta \sim 0.7$ Pa.s) with triethylenetetramine (TETA) hardener (Epofix, Struers) was used. The stochiometric ratio was 100:12 (w/w) epoxy resin:hardener.

### 2.2 Sample preparation

The as-received CNF powder contained many large clumps with sizes in the mm range. Sieving as reported elsewhere [13, 14] was used as a standard debulking method. The CNF powder was manually pushed through a 0.5mm sieve. Carbon nanofibres were then dispersed in acetone in a concentration of 0.3 vol % using tip ultrasonication in pulse mode (30 pulses per minute) combined with mechanical mixing. The epoxy resin was diluted in acetone (1:4 in volume) and then added to the CNF solution. The mixture was again submitted to ultrasonication in pulse mode (30 pulses per minute) combined with mechanical mixing for 4 hours. Solvent was removed by heating under continuous mixing. A masterbatch containing 10 wt % of CNF was obtained in that way. Composites with different CNF loadings ranging from 0.066 to 9 wt % were prepared by dilution of the masterbatch with the right amount of resin. The mixture was mixed under vaccum with a Thinky mixer at 2000 rpm. The hardener was added, the mixture was again mixed under vacuum, then cast into a mold and cured at 120 °C for two hours. Disk-shaped samples with diameter 25 mm were obtained and polished down to a thickness around 1 mm.

**2.3 Impedance spectroscopy**

The electrical properties were investigated using a dielectric analyzer (TA Instrument DEA 2970) in ceramic parallel plate mode. All experiments were performed at room temperature and at testing frequencies ranging from 1 to $10^5$ Hz. Nitrogen gas was used to provide an inert environment at a flowing rate of 500 ml/min. The sample was placed between two gold electrodes, a load (300 N) was applied and the thickness was measured with the built-in LVDT (linear voltage-displacement transducer) with a precision of 1 µm. A low amplitude sinusoidal voltage ($V_{applied}$) was applied and the current ($I_{measured}$) through the sample was measured. The AC conductivity is given by

$$\sigma_{AC} = \frac{I_{measured}}{V_{applied}} (\cos \delta) \frac{e}{A}$$

with $\delta$ being the phase angle shift, $e$ being the thickness and $A$ the surface area of the sample.

## 3. RESULTS AND DISCUSSION

### 3.1 Microstructure

Light transmission optical micrographs at low magnification permitted the observation of the gradual formation of a complex network and the global homogeneity of the sample. As the weight fraction increased, the aggregation appeared more refined and highly structured, as can be seen in Figure 2a. At a higher magnification (Figure 2b), the interconnected network of CNF is revealed. It can be observed that CNF have different lengths ranging from 1 to 10 microns. As previously reported [13], when using mechanical mixing for dispersing CNF in epoxy, isolated aggregates with size around 30 microns remained. This size almost corresponds to the length of the CNF, as a consequence, it is difficult to break these aggregates without CNF breakage and length reduction. The effect of fibre breakage is to shift the CNF critical fraction at which the composite becomes conductor to higher values. In this work, the effect of fibre breakage is negligible compared to the beneficial effect of desaggregation as the critical fraction is reduced from 1 to less than 0.1 wt %. The better the dispersion the lower the critical fraction.

### 3.2 Electrical conductivity

The electrical conductivity, $\sigma$, as a function of frequency is plotted in Figure 3 for the composites at different CNF loadings. For each loading, at least three samples were tested and the reproducibility is good. As expected, the standard deviation is higher in the transition region. The unfilled epoxy resin showed a typical insulating behaviour with a frequency-dependent conductivity. The composite with 0.066 wt % has the same behaviour as the matrix with a shift of its conductivity about two orders of magnitude compared to the virgin resin. At 0.1 wt % loading, the behaviour of the composite became semi-conducting. The conductivity is independent of the frequency until an onset frequency, noted as $f_0$, at which it increased with frequency. From 0.1 to 1.2 wt %, the same behaviour was observed with a shift of $f_0$ to higher frequencies. This kind of behaviour was previously reported for carbon black composites [15, 16] and is also described by universal scaling laws [17]. A model was proposed by Connor et al. [15] based on the fact that an additional contribution to conductivity comes from electrons in finite-size clusters with fractal nature that can be scanned at higher frequencies (higher than $f_0$). Percolation theory predicts power law scaling of conductivity, $\sigma$, and dielectric constant, $\varepsilon$, with frequency $\omega$ close to the transition region; $\sigma \propto \omega^x$ and $\varepsilon \propto \omega^{-y}$. Different mechanisms [18] (intercluster polarization (x=0.72, y=0.28), anomalous diffusion inside cluster (x=0.58, y=0.42)) have been proposed to explain the conduction behaviour. In our case, x takes values ranging from 0.45 to 0.65, and y takes values ranging from 0.17 to 0.25, for $p$ in the range 0.23 to 0.8 wt %, and these values do not appear to correspond closely to the defined mechanisms. On the other hand, $\log \sigma$ is linear with $p^{-1/3}$, as can be seen in Figure 4, showing the possible preponderance of the tunneling conduction mechanism [15]. For higher loadings ($\geq$ 1.2 wt %), the onset frequency disappeared and the composite presented a conducting behaviour over the whole domain of frequency studied. The AC conductivity at 1 Hz is plotted as a function of wt % of CNF in Figure 5. The electrical conductivity increased by seven orders of magnitude with the addition of 1.2 wt % of CNF into the epoxy resin. The data corresponding to the vicinity of the transition, that is from 0.1 to 1.2 wt %, was fitted to the universal scaling law [19, 20]:

$$\sigma = A\,(p - p_c)^t \quad (\text{for } p < p_c) \qquad (1)$$

with $\sigma$, the electrical conductivity; A, a constant; $p$, the weight fraction of CNF; $p_c$, the critical weight fraction of CNF, and $t$ the critical exponent related to the dimensionality of the system. It was found that the critical weight fraction is 0.064 wt % and the exponent is 1.78, which is close to that of a three dimensional random system ($t = 1.94$) [20]. This critical weight fraction corresponds to the formation of a tunneling conductive network. This is a network of aggregates that are close enough from each other (~1 nm) to allow for tunneling conduction between them. As an illustration of this aspect, the case of the composite with 0.066 wt % of CNF (thus $p > p_c$) has an insulating behaviour with a conductivity two orders of magnitude higher than that of the unfilled resin. It is also interesting to observe that the conductivity increased by two orders of magnitude after the transition and reached $5.5 \times 10^{-5}$ S/cm at 9 wt %.

**3.3 A tunneling contact model**

After the transition, for CNF loadings higher than 1.2 wt %, the electrical conductivity did not change significantly and was around $10^{-6}$ S/cm, before increasing again by about two orders of magnitude. The value of the conductivity right after the transition is lower than expected, considering that the conductivity of the pure CNF powder at 10 vol % is around 1 S/cm as reported by the supplier and as measured by our own means. Similar conclusions were drawn in a previous study [21]. In that study, the electrical conductivity of a network of multiwall carbon nanotubes (MWNT) was measured as a function of the volume fraction of MWNT. The electrical conductivity of MWNT/epoxy composites prepared with the same MWNT was also measured. The comparison of the results showed up to three orders of magnitude of difference of the extrapolated conductivity of the 100 vol % MWNT material. This discrepancy led to the conjecture that a thin film of epoxy was present at the contact between MWNT. When the thickness of that insulating film is sufficiently small (~1 nm), quantum mechanics laws state that the probability to find an electron on the other side of the film is non-zero and the electrons can cross this insulating barrier by "tunnel effect". We propose to evaluate the thickness of the epoxy film. We model the composite as a stacking of layers. The resistance of one layer is that of a network of parallel resistors. Each resistor has the same resistance, noted as $R_{contact}$, and represents the contact between the CNF with an epoxy film in their vicinity along with the segment of CNF between contacts. The thickness of one layer is assumed to be equal to the distance between contacts, noted as $\lambda$. A composite sample of thickness $e$ is a stacking of $e/\lambda$ layers (the layers resistances are in series). The electrical resistance of the composite is $R = \frac{e}{\lambda} \frac{R_{contact}}{N}$ with $N$ the number of contacts (in parallel) in one layer. Considering a composite sample of surface area $S$, the electrical conductivity is thus $\sigma = \frac{1}{R}\frac{e}{S} = \frac{\lambda}{e}\frac{n\lambda S}{R_{contact}}\frac{e}{S} = \frac{n\lambda^2}{R_{contact}}$ with $n$ being the number of contacts per unit volume in a three dimensional random fibre network which is given by [22]

$$n = \frac{4\phi^2}{\pi d^3} \qquad (2)$$

with $\phi$ the volume fraction of fibres and $d$ their diameter. The distance between contacts in a three dimensional random fibre network is given by [22]

$$\lambda = \frac{\pi d}{8\phi} \qquad (3)$$

Finally, the electrical conductivity of the composite is simply

$$\sigma = \frac{\pi}{16d} \frac{1}{R_{contact}} \tag{4}$$

At least two assumptions are made. In each layer, the network is perfectly 3D random and all the contacts participate in the conduction (we do not take into account aggregation and dead-end branches: far from the transition, the latter may have a limited effect). All resistors have the same resistance, that is we do not take into account the distribution in diameter of the CNF or variation of epoxy film thickness. The resistance of a contact is the sum of $R_{epoxyl}$, the resistance of an epoxy film with thickness $s$ and surface area equal to that of a CNF-CNF contact, and of $R_{CNFp}$, the resistance of the CNF portion between two contacts.

$$R_{contact} = R_{CNFp} + R_{epoxyl} \tag{5}$$

A schematic illustration of the model is given in Figure 6. We do not take into account the complexity of the contact surface and we approximate it by a square of side equal to the diameter of a CNF; $S_{contact} = d^2$. This means that all CNF are assumed to cross each other under a right angle and thickness variation along the section is neglected. Hence,

$$R_{epoxyl} = \rho_{tunnel} \frac{1}{d^2} \tag{6}$$

with $\rho_{tunnel}$ the tunneling resistivity of the epoxy film (in $\Omega.cm^2$). Even if we know that the resistance between two contacts, that is the resistance of a portion of CNF, is negligible compared to that of a contact (this makes sense, the conductivity along a CNF is $10^3$ S/cm compared to the bulk conductivity of a network of CNF at 10 vol % around 1

S/cm), we will include this term in order to keep the generality of the model. As a consequence, the resistance of the CNF portion between two contacts is

$$R_{CNFp} = \frac{1}{\sigma_{CNF}} \frac{\lambda}{\pi d^2/4} = \frac{1}{\sigma_{CNF}} \frac{1}{2d\phi} \tag{7}$$

The combination of equations (4), (5), (6) and (7) led to the following expression for the electrical conductivity of the composite

$$\sigma = \frac{\pi}{16d} \frac{1}{\rho_{tunnel}\frac{1}{d^2} + \frac{1}{2d\phi\sigma_{CNF}}} \tag{8}$$

The volume fraction of CNF was given by $\phi = \left(1 + \frac{\rho_{CNF}}{\rho_{epoxy}} \frac{p_{epoxy}}{p_{CNF}}\right)^{-1}$ with $\rho_{CNF}$ = 2 g.cm$^{-3}$ and $\rho_{epoxy}$ = 1.1 g.cm$^{-3}$ the densities of CNF and epoxy, and $p_{CNF}$ and $p_{epoxy}$ the weight fractions of CNF and epoxy. An important research effort [23-26] was dedicated to the calculation of the tunneling resistivity based on quantum considerations. For low voltages, the Holm-Kirschstein [24, 25] equation gives the tunnel resistivity in $\Omega.cm^2$

$$\rho_{tunnel} = \frac{10^{-22}}{2} \frac{A^2}{1+AB} e^{AB} \tag{9}$$

with $A = 7.32 \times 10^5 \left(s - \frac{7.2}{\Phi}\right)$ and $B = 1.265 \times 10^{-6} \sqrt{\Phi - \frac{10}{s\varepsilon_r}}$

$s$ being the film thickness in Å, $\Phi$ the work function of the metal in eV and $\varepsilon_r$ the relative permittivity of the film material. Simmons [26] made refined calculations taking into account the true shape of the potential and not the approximate parabolic form assumed by Holm et al. [24, 25]. He derived the following expression of the tunneling current density, $J$, in A.cm$^2$

$$J = \frac{6.2 \times 10^{10}}{s^2} \left\{ \varphi \times \exp(-1.025 s \varphi^{1/2}) - (\varphi + V) \exp\left(-1.025 s (\varphi + V)^{1/2}\right) \right\} \quad (10)$$

where $\varphi = \Phi - \left(\dfrac{V}{2s}\right)(s_1 + s_2) - \left[\dfrac{5.75}{\varepsilon_r (s_2 - s_1)}\right] \ln\left(\dfrac{s_2 (s - s_1)}{s_1 (s - s_2)}\right)$

and

$$s_1 = \frac{6}{\varepsilon_r \Phi}$$

$$s_2 = s\left(1 - \frac{46}{3\Phi \varepsilon_r s + 20 - 2V \varepsilon_r s}\right) + \frac{6}{\varepsilon_r \Phi}$$

with the film thickness $s$ expressed in Å and the work function $\Phi$ in eV. The electroaffinity of the film material should be taken into account, but in the case of an insulator, it is assumed that it may be negligible compared to the work function. The voltage, $V$, is given by considering one electron carrying the elementary charge $+e$ passing into the film of thickness $s$, surface area $S_{contact}$ and capacitance $C$ $V = e/C = e/(\varepsilon_r \varepsilon_0 S/s)$ with $\varepsilon_r$ and $\varepsilon_0 \sim 8.85 \times 10^{-12}$ F/m, the permittivity of respectively the film and vacuum. The tunneling resistivity is then obtained, $\rho_{tunnel} = V/J$. The tunnel resistivity was calculated using the relative permittivity of the epoxy $\varepsilon_r \sim 2.65$, the work function of graphite [27] $\Phi = 4.62$ eV and an average CNF diameter $d = 100$ nm. We assumed that the film thickness varies with the volume fraction of CNF following a power law dependence; $s = K\phi^\beta$. The experimental data were fitted to equation (8) with $K$ and $\beta$ as free parameters using the two different models for the tunneling resistivity;

Holm-Kirschstein and Simmons. The experimental data along with the fitting curves are plotted in Figure 7. The fitting curves are superimposed and the values of the fitting parameters are quite similar. The Simmons model is a refined version of Holm's model, the difference is the potential used to calculate the tunnel current, in Holm's model, there is a more crude approximation. The thickness of the epoxy film is plotted in Figure 8 using the values of these fitting parameters and the assumed power law dependence with the volume fraction of CNF. It was found that the film thickness is in the range of 12-20 Å and decreases when the volume fraction of CNF increases. This thickness range corresponds to the size of a few epoxy molecules [28]. The difference in thickness between the two models was less than 2 Å and the trend was quite similar.

## 4. CONCLUSIONS

In this study, carbon nanofibre-based composites were prepared using a method combining ultrasonication and mechanical mixing. The study of their electrical properties by impedance spectroscopy revealed a low critical weight fraction, $p_c$ = 0.064 wt %, corresponding to the formation of a tunneling conductive network. The transition region was quite large, spanning about one order of magnitude, from 0.1 to 1.2 wt %. In that region, the composites presented a conducting behaviour until an onset frequency at which their conductivity increased with frequency, that is an insulating behaviour. The onset frequency increased with CNF weight fraction and disappeared for $p \geq 1.2$ wt %. Far from the transition, the conductivity increased by two orders of magnitude. A simple model, based on the CNF-CNF contact inside the matrix and assuming the presence of an epoxy film in the vicinity of the contact, was proposed in order to evaluate the thickness of this film. It appears that the thickness corresponds to the size of a few epoxy molecules and decreased with CNF weight fraction. The decrease of the epoxy film may explain the two orders of magnitude increase in the conductivity observed far from the transition.

## ACKNOWLEDGEMENT

Financial support from the Natural Sciences and Engineering Research Council of Canada (NSERC) is appreciated.

**LIST OF CAPTIONS**

Figure 1 – Scanning electron micrograph of the carbon nanofibres.

Figure 2 – Light transmission optical micrographs of the composites at different loadings of CNF ranging from 0.066 to 1.2 wt %. a) magnification 50X b) magnification 1600X.

Figure 3 – Electrical conductivity of the composites at different CNF loadings as a function of frequency.

Figure 4 – Logarithm of the conductivity as a function of $p^{-1/3}$. Solid line is a linear fit to the data.

Figure 5 - Electrical conductivity of the composites at 1 Hz as a function of weight fraction of CNF. The solid line is a fit to the universal scaling law. (Inset: logarithm of conductivity as a function of logarithm of reduced weight fraction and fit to a line).

Figure 6 – Sketch of the model adapted from [21].

Figure 7 – Electrical conductivity as a function of volume fraction of CNF. The solid line is a fit to equation (8).

Figure 8 – Thickness of the epoxy film as a function of volume fraction of CNF.

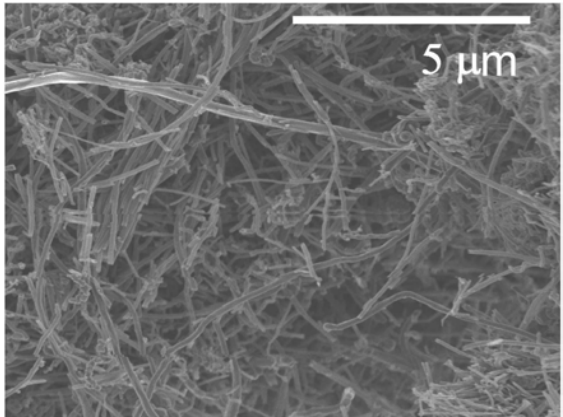
Figure 1 – Scanning electron micrograph of the carbon nanofibres.

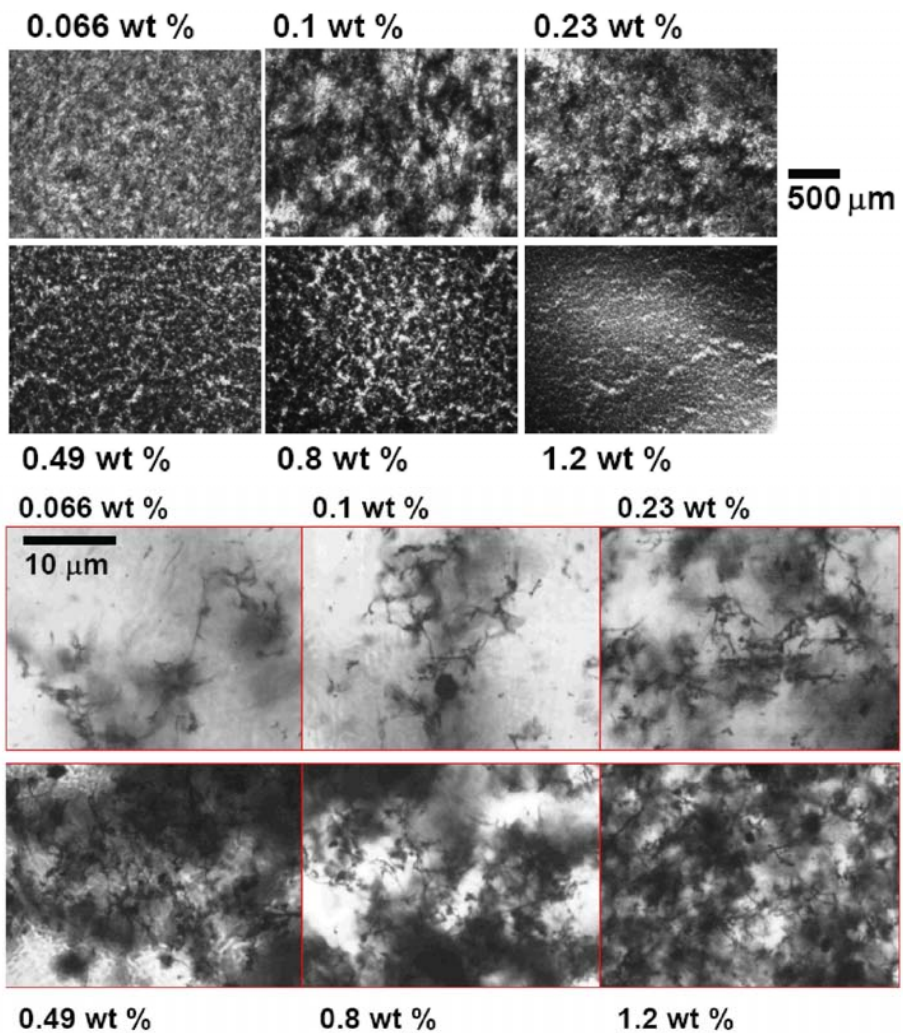
Figure 2 – Light transmission optical micrographs of the composites at different loadings of CNF ranging from 0.066 to 1.2 wt %. a) magnification 50X b) magnification 1600X.

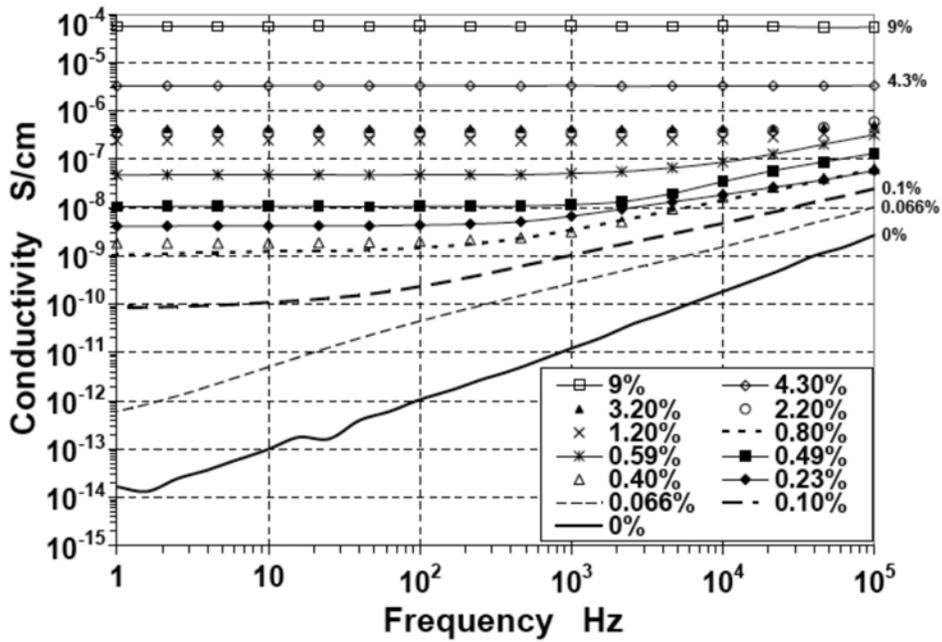

Figure 3 – Electrical conductivity of the composites at different CNF loadings as a function of frequency.

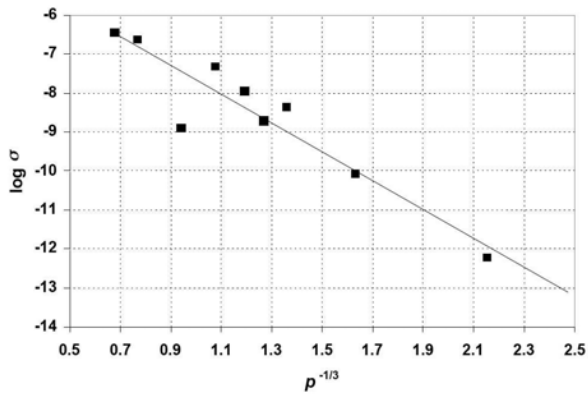

Figure 4 – Logarithm of the conductivity as a function of $p^{-1/3}$. Solid line is a linear fit to the data.

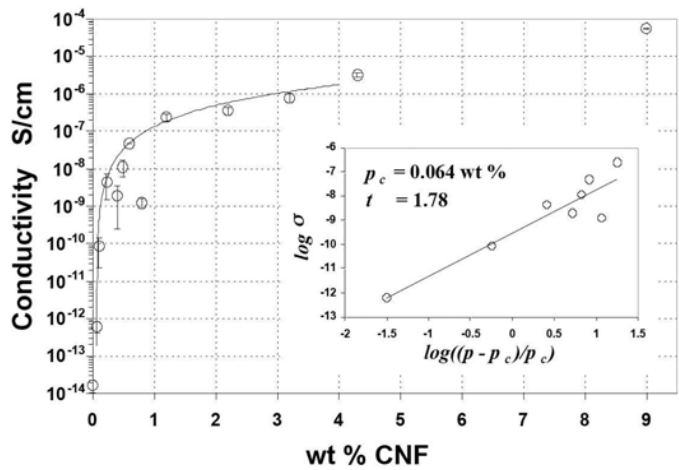

Figure 5 - Electrical conductivity of the composites at 1 Hz as a function of weight fraction of CNF. The solid line is a fit to the universal scaling law. (Inset: logarithm of conductivity as a function of logarithm of reduced weight fraction and fit to a line).

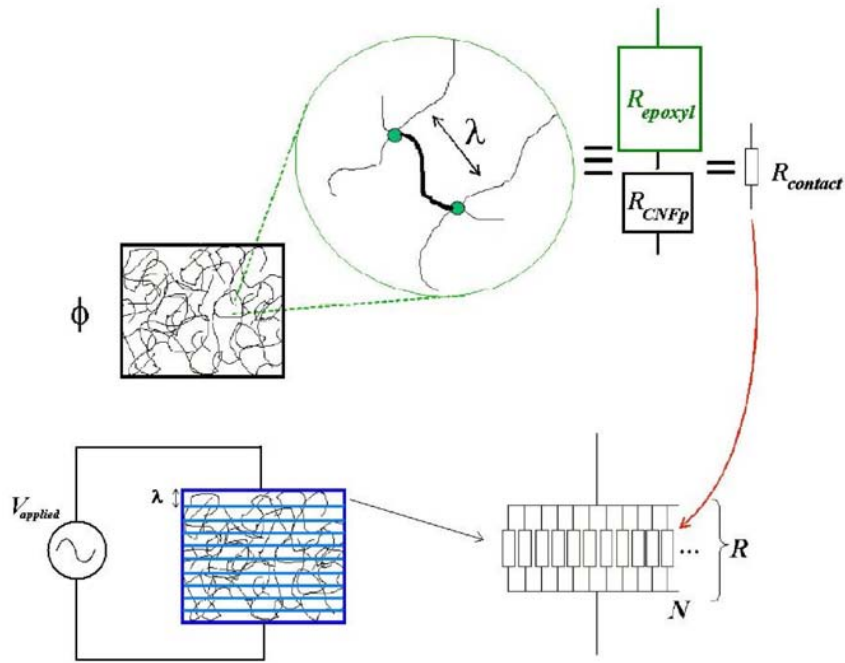

Figure 6 – Sketch of the model adapted from [21].

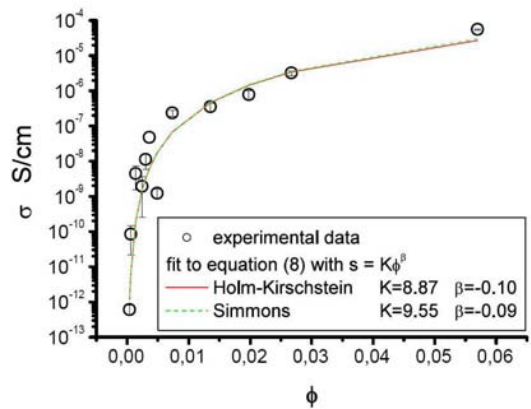

Figure 7 – Electrical conductivity as a function of volume fraction of CNF. The solid line is a fit to equation (8).

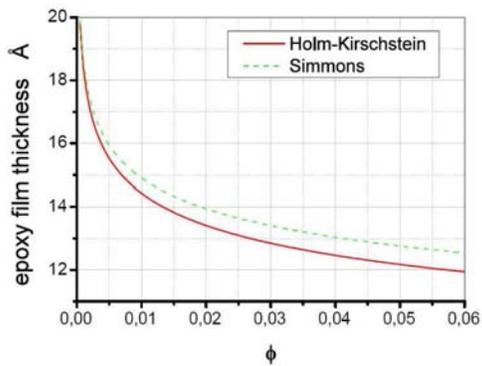

Figure 8 – Thickness of the epoxy film as a function of volume fraction of CNF.